# Rules of plastic strain-induced phase transformations and nanostructure evolution under high-pressure and severe plastic flow

*Feng Lin[1,*], Valery I. Levitas[1,2,3,*], Krishan K. Pandey[4], Sorb Yesudhas[1], and Changyong Park[5]*

[1]Department of Aerospace Engineering, Iowa State University, Ames, IA 50011, USA
[2]Departments of Mechanical Engineering, Iowa State University, Ames, IA 50011, USA
[3]Ames Laboratory, Division of Materials Science and Engineering, Ames, IA 50011, USA
[4]High Pressure & Synchrotron Radiation Physics Division, Bhabha Atomic Research Centre, Mumbai 400085, India
[5]HPCAT, X-ray Science Division, Argonne National Laboratory, Argonne, Illinois 60439, USA

*Abstract.* Rough diamond anvils (rough-DA) are introduced to intensify all occurring processes during an in-situ study of heterogeneous compression of strongly pre-deformed Zr in diamond anvil cell (DAC). Crystallite size and dislocation density of Zr are getting pressure-, plastic strain tensor- and strain-path-independent during α-ω phase transformation (PT) and depend solely on the volume fraction of ω-Zr. Rough-DA produce a steady nanostructure in α-Zr with lower crystallite size and larger dislocation density than smooth-DA, leading to a two-time reduction in a minimum pressure for α-ω PT to a record value 0.67 GPa. The kinetics of strain-induced PT unexpectedly depends on time.

*Introduction.* Processes that combine severe plastic strain and PTs under high pressure are widespread in manufacturing, synthesis of nanostructured materials, and geophysics-related problems [1-13]. Plastic strain drastically reduces the PT pressure by up to one-two order of magnitude [3, 5, 6], lead to new nanostructured phases, and substitute time-controlled kinetics with plastic strain-controlled kinetics [7-9]. Four-scale theory and simulations [7, 8] are developed to explain strain-induced PTs (which are completely different from traditional pressure or stress-induced PTs). However, they are still in their infancy, and new experimental and theoretical breakthroughs are required. In particular, the mutual effect between the nanostructure evolution and strain-induced PTs was not studied in situ at all. The main problem is that these processes depend on pressure, five components of the plastic strain tensor $\boldsymbol{\varepsilon}_p$, and the entire strain path $\boldsymbol{\varepsilon}_p^{path}$, producing numerous combinations of independent parameters with little hope of fully comprehending. For example, different combinations of compression and shear resulting in the same final $\boldsymbol{\varepsilon}_p$ lead to different stresses, nanostructures, and volume fractions of the high-pressure phase. Other problems include very heterogeneous stress-strain fields in all typical deformation-transformation processes and limited in-situ measurement capabilities. For example, in high-pressure torsion (used for grain refinement and producing nanostructured materials [1, 10, 14, 15]), all measurements are performed postmortem after pressure release and further treatment during sample preparation for mechanical and structural studies. Pressure is determined as force per unit area, while local pressure can be several times higher. For compression in DAC or torsion in rotational DAC, radial pressure distribution in phases and volume fraction distributions were recently determined only in [9]. Distributions of plastic strain and structural parameters were never measured.

Here, we compress severely pre-deformed Zr in DAC and perform the first in-situ study of the coupled strain-induced α (hcp)-ω (simple hexagonal) Zr PT kinetics and nanostructure evolution in terms of crystallite size and dislocation density with synchrotron x-ray diffraction and found new general rules and unexpected phenomena. Zr is an important engineering and structural material with various applications in the nuclear, biomedical, and aerospace industries. Ambient α-Zr sample is heavily deformed by multiple rolling until reaching steady hardness, crystallite size and dislocation density, which greatly simplifies studies by excluding significant nanostructure evolution during compression in DAC before PT. We introduced rough-DA, whose culets are roughly polished to increase friction with a sample, intensify plastic straining, strain-induced PTs and structural changes, and enrich classes of the plastic strain tensors $\boldsymbol{\varepsilon}_p$ and entire strain paths $\boldsymbol{\varepsilon}_p^{path}$ that material undergoes. The following unexpected and informative rules and phenomena are found. With rough-DA, a record minimum pressure for strain-induced α-ω PT $p_\varepsilon^d$ =0.67 GPa is obtained, which is two times lower than with smooth anvils (1.36 GPa), and both are independent of strain $\boldsymbol{\varepsilon}_p$ and strain path $\boldsymbol{\varepsilon}_p^{path}$. The difference is related to higher steady dislocation density and lower crystallite size before PT achieved with rough-DA. This PT pressure is also 9.0 times lower than that under hydrostatic loading (6.0 GPa) and 5.1 times lower than the phase equilibrium pressure of 3.4 GPa [16]. The minimum PT pressure $p_\varepsilon^d$ reduces with the reduction in the crystallite/grain size, opposite to the existing theoretical prediction [7]; a more advanced theory explains this result below. Surprisingly, dislocation density and crystallite size in ω-Zr during the PT are found to be independent of pressure $p$, strain $\boldsymbol{\varepsilon}_p$, and strain path $\boldsymbol{\varepsilon}_p^{path}$, and solely are functions of ω-Zr volume fraction $c$ for both smooth and rough-DA and different nanostructure before PT. Similar rule was found for nanostructure in α-Zr with rough-DA, but with some scatter. In addition, an unexpected time-dependence of PT kinetics is revealed with rough-DA, which confronts the conventional view that strain-induced PTs do not occur at fixed plastic strain. To rationalize it, our theory is advanced accordingly. While with smooth-DA, strain-controlled part of the PT kinetics is the first order (as expected), with rough-DA, it is a zero-order kinetics. Experimental methods and details are presented in Supplemental Materials [17].

*Record minimum pressure for strain-induced α-ω PT.* Compression steps are named after the corresponding peak pressure at the culet center for convenience. During compression with rough-DA, ω-Zr diffraction peaks start to be observed at $p_\varepsilon^d$ =0.67 GPa at the sample center (Fig. 1a and S1). This is a record low pressure for α-ω Zr PT, which is 9.0 times lower than that under hydrostatic loading ($p_h^d$ =6.0 GPa) from this study, 5.1 times lower than the phase equilibrium pressure of 3.4 GPa [16], and 2 times lower than $p_\varepsilon^d$ =1.36 GPa obtained here with smooth anvils. At the culet edge at 2 GPa step, $c$=0.05 at 0.74 GPa (Fig. 1b), meaning $p_\varepsilon^d$ at the edge is practically identical to that at the center. The same is true for smooth diamonds (Fig. S2), for which, due to higher PT pressure, we have more points with $p=p_\varepsilon^d$ at different positions. While we cannot measure $\boldsymbol{\varepsilon}_p$ field, it is known from the finite element simulations of the processes in DAC [18-20] and Fig. S3 that for different material positions and compression stages, $\boldsymbol{\varepsilon}_p$ and $\boldsymbol{\varepsilon}_p^{path}$ vary substantially. In particular, there is no shear strain at the symmetry axis, and shear increases with increasing radius. This indicates that for strongly pre-deformed α-Zr, $p_\varepsilon^d$ is independent of $\boldsymbol{\varepsilon}_p$, $\boldsymbol{\varepsilon}_p^{path}$, and pressure-strain path since they are very different at the center and edge. This is the main and very informative rule we found for strain-induced PTs under pressure, which drastically simplifies the theory.

We explain the difference in $p_\varepsilon^d$ by different steady nanostructures at the initiation of PT produced by different DA. At the initiation of PT with rough-DA, $d_\alpha$ =46 nm and $\rho_\alpha$= 1.68×10$^{15}$m$^{-2}$, while with smooth anvils, $d_\alpha$=66 nm and $\rho_\alpha$= 1.22×10$^{15}$m$^{-2}$, both represent steady value (state) independent of radii and, consequently, $\boldsymbol{\varepsilon}_p$, $\boldsymbol{\varepsilon}_p^{path}$, and pressure-strain path. Rough-DA produce more refined crystallite size and higher dislocation density in α-Zr, resulting in two times lower pressure for initiating strain-induced PT $p_\varepsilon^d$. Since for annealed α-Zr with micron-size grains, $p_\varepsilon^d = 2.3\ GPa$ [9], a general trend is that $p_\varepsilon^d$ reduces with reduction in $d_\alpha$ (opposite to the initial theoretical prediction in [7]) and increase in $\rho_\alpha$. This dependence also explains the large scatter in pressures for α-ω PT during high-pressure torsion [4, 9, 21-24]: since grain size and dislocation density at the initiation of PT were not measured, they may vary substantially and cause the difference in PT pressure.

Let us explain the obtained result and resolve the contradiction with analytical models. As suggested in analytical [7] and phase field models [25, 26], plastic strain-induced PT occurs by nucleation at the tip of a dislocation pileup as the strongest possible stress concentrator. All components of stress tensor $\boldsymbol{\sigma}$ at the tip of dislocation pileup, modeled as a superdislocation, are $\sigma \sim \tau l$, where $\tau$ is the applied shear stress limited by the yield strength in shear $\tau_y$, and $l$ is the length of the dislocation pileup [27]. The higher the dislocation density is, the higher the probability of the appearance of dislocation pileups with a larger number of dislocations is. This trivially explains reducing the minimum pressure for the strain-induced PT with increasing dislocation density. However, since *l* is traditionally limited by the fraction of the grain size (e.g., half of the grain size), the main conclusion in [7] was that the greater grain size is, the stronger the stress concentrator and consequently reduction in the PT pressure, which opposes to what we found in the current experiments. Later phase field [25, 26], molecular dynamics [28], and concurrent atomistic-continuum simulations [29] allow us to resolve the problem, at least qualitatively. In contrast to the analytical solution utilized in [7], *l* is not related to the grain size *d* since most dislocations are localized at the grain boundary producing a step (superdislocation, Fig. S4) with effective length *l=Nb<<d*, where *b* is the magnitude of the Burgers vector and *N* is the number of dislocations in a pileup. At the same time, $\tau = \tau_y$ increases with the decrease in *d* according to the Hall-Petch relationship $\tau_y = \tau_0 + kd^{-0.5}$, where $\tau_0$ and *k* are material parameters. That is why the minimum pressure for the strain-induced PT decreases with decreasing crystallite size. Such a trend may not be valid for grain size in the inverse Hall-Petch effect region, which needs to be checked in the future.

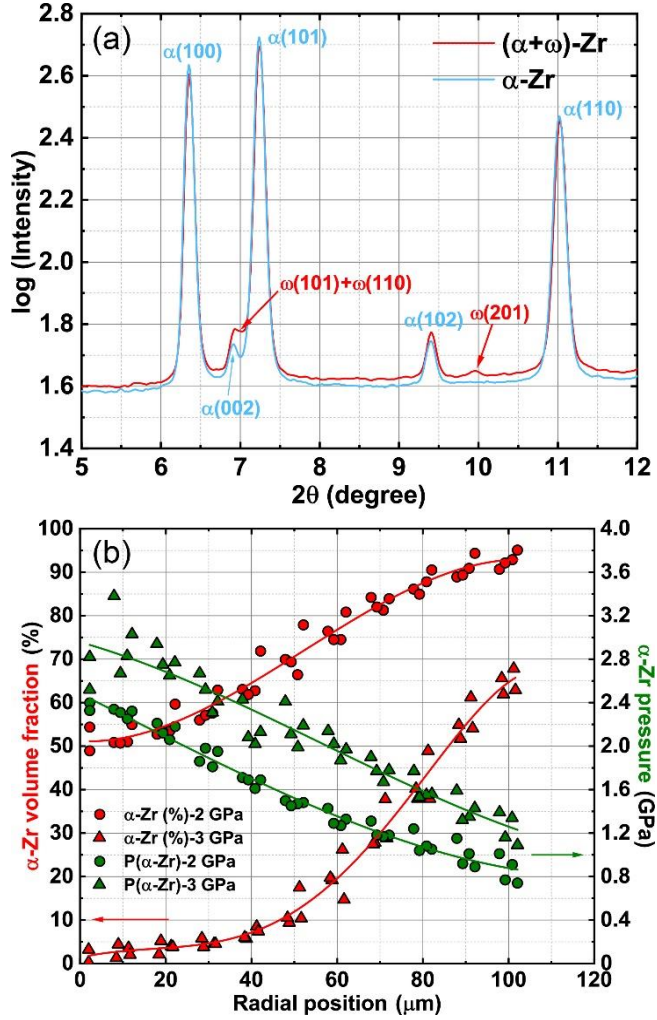

**Fig. 1.** (a) X-ray diffraction patterns from the sample center obtained with rough-DA demonstrating pure α-Zr diffraction peaks (blue) at $p$=0.49 GPa and appearance of ω-Zr peaks at the record minimum pressure $p_\varepsilon^d = 0.67$ GPa (red). (b) Distribution of volume fraction of α-Zr and pressure in α-Zr at 2 and 3 GPa steps. At the edge of 2 GPa step, ~5% ω-Zr is observed at the pressure of 0.74 GPa. Note that the uncertainty of volume fraction, pressure, crystallite size, and dislocation density from the Rietveld refinement are smaller than the symbol.

*Nanostructure evolution during PT.* The pressure, nanostructure, and ω-Zr volume fraction are scrutinized along two perpendicular diameters for 2 and 3 GPa steps (Fig. 1b and Fig. 2). At the edge of 2 GPa step, where ω-Zr volume fraction is tiny (~5%), the crystallite size of ω-Zr is as small as ~10 nm. With increasing volume fraction towards the culet center due to higher pressure, crystallites of ω-Zr grow and reach a similar size as the original α-Zr (~50 nm) at the culet center at 3 GPa step, where α-Zr is nearly fully transformed. The crystallite size of α-Zr remains constant during PT until volume fraction of α-Zr is below ~50% and then gradually decreases with diminishing α-Zr. The crystallite size of α-Zr slightly increases at the culet center at 2 GPa step due to statistical effect since smaller α-Zr grains are easier to be transformed (see [17] for discussion). Dislocation density variation generally follows the inverse trend of crystallite size from Eq. (S1). It is worth noting that at the edge of 2 GPa step, where ω-Zr starts to emerge, the dislocation density of α and ω-Zr are very close (Fig. 2b), indicating that ω-Zr nuclei inherit the dislocations from α-Zr without modification.

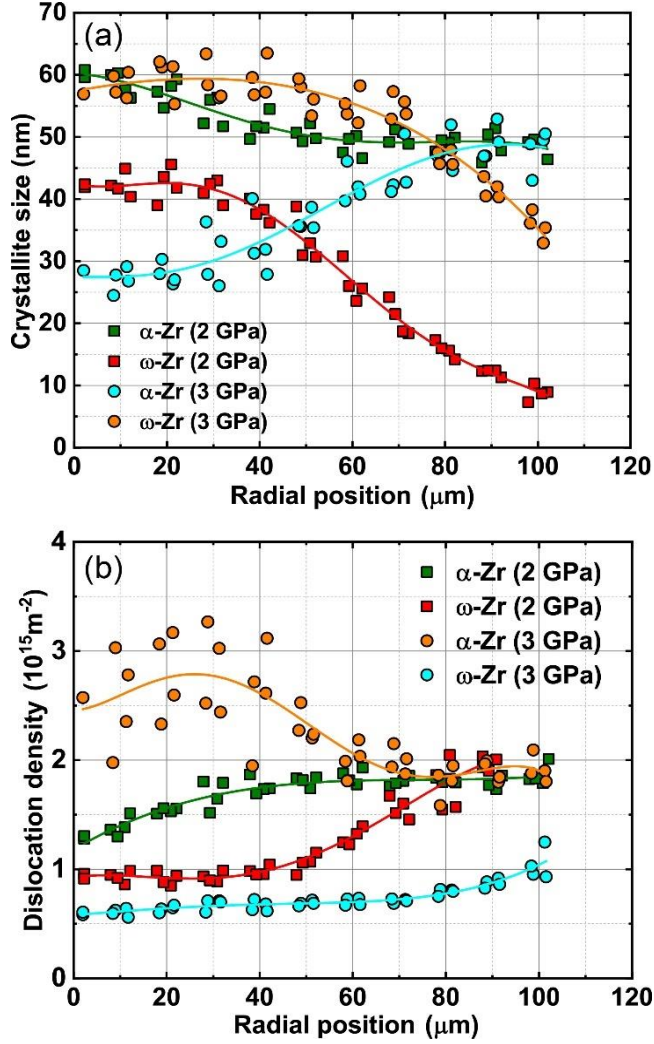

**Fig. 2.** Radial distributions of (a) the crystallite size; (b) the dislocation density; in α-Zr and ω-Zr at 2 and 3 GPa steps with rough-DA.

To find the main rule, crystallite size and dislocation density in α-Zr and ω-Zr for all cases are plotted as a function of ω-Zr volume fraction c (Fig. 3). Crystallite size and dislocation density of ω-Zr overlaps for 2 GPa around the center and 3 GPa around edge within the shared volume fraction region (Fig. 3), and data from all radial positions for both loading steps (i.e., for variety of $p$, $\varepsilon_p$, and $\varepsilon_p^{path}$) can be described by a single curve as a function of just one argument, $c$. For compression with smooth anvils, which gives different $p$, $\varepsilon_p$, $\varepsilon_p^{path}$, and initial microstructure of α-Zr ($d_\alpha = 66$ nm and $\rho_\alpha = 1.22 \times 10^{15}$ m$^{-2}$), the crystallite size and dislocation density of ω-Zr follow the same curves. Thus, one more surprising rule is found for ω-Zr: the existence of the unique curves $d_\omega(c)$ and $\rho_\omega(c)$ during α-ω PT, which are independent of pressure, $\varepsilon_p$, and $\varepsilon_p^{path}$, and initial nanostructure of α-Zr (Fig. 3a, b). This greatly simplifies the theory and controls the nanostructure evolution and PT processes. For α-Zr and rough DA, there is a similar but weaker regularity (Fig. 3c, d) with some scatter for $0.4<c<0.5$ from the statistical effect [17].

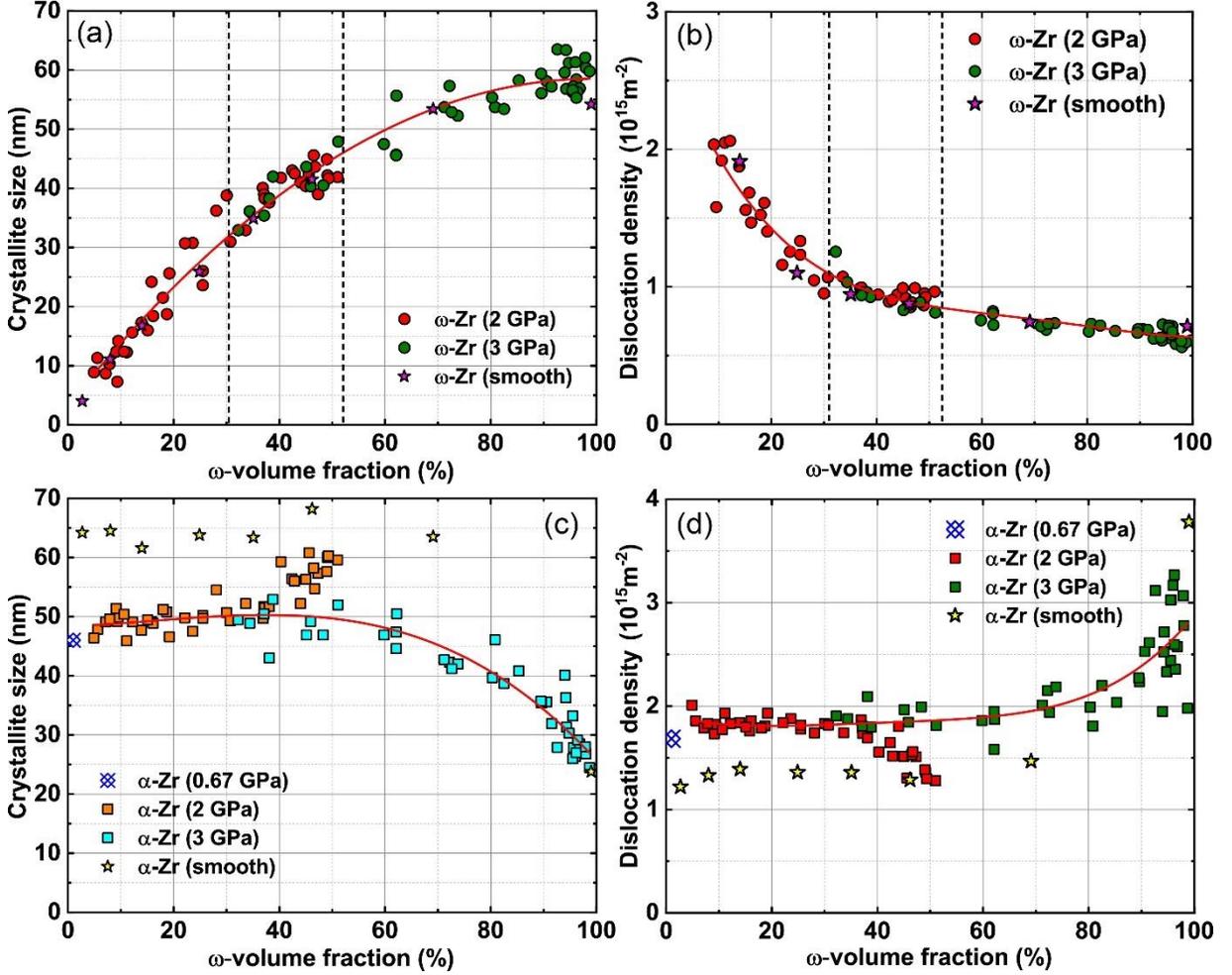

**Fig. 3**. (a) Crystallite size and (b) dislocation density in ω-Zr versus volume fraction of ω-Zr at 2 and 3 GPa steps with rough-DA and smooth-DA. For rough-DA, points from 2 and 3 GPa steps overlap within dash lines. (c) Crystallite size and (d) dislocation density in α-Zr versus volume fraction of ω-Zr at 2 and 3 GPa steps with rough-DA and smooth-DA. Results in (a) and (b) represent a remarkable rule for ω-Zr: the existence of the unique curves for the crystallite size and dislocation density solely depending on $c$ during α-ω PT, which are independent of pressure, $\varepsilon_p$ and $\varepsilon_p^{path}$, the same for rough- and smooth-DA and different nanostructure of α-Zr at the initiation of PT. A similar rule is valid for α-Zr in (c) and (d) for rough-DA, except for region $0.4<c<0.5$, where some scatter is observed.

*Time dependence of strain-induced PT and zero-order PT kinetics.* Distributions of pressure in phases and volume fraction c of ω-Zr at 2 and 3 GPa steps are presented in Figure 1b. Strain-induced PT kinetic equation derived based on nanoscale mechanisms [7] with neglected reverse

PT is:

$$\frac{dc}{dq} = k \frac{B(1-c)^a}{B(1-c)+c} \left(\frac{p_\alpha(q)-p_\varepsilon^d}{p_h^d-p_\varepsilon^d}\right) \quad \text{for} \quad p_\alpha > p_\varepsilon^d. \tag{1}$$

Here $q$ is the accumulated plastic strain, $p_\alpha(q)$ is the pressure in α-Zr - q loading path; $B = \left(\frac{\tau_y^\omega}{\tau_y^\alpha}\right)^l$; $k$, $a$, and $l$ are material parameters. Although plastic strain tensor at arbitrary $r$ is unknown in experiments, material near the symmetry axis undergoes macroscopically uniaxial compression for which $q = ln(h_0/h)$. Through numerical integration of Eq. (1) with spline interpolation of p-q path using experiment points, $c$ can be expressed as a function of $I = \int_{q_0}^{q} (p_\alpha(q) - p_\varepsilon^d) \, dq$, where $q_0$ is the accumulated plastic strain at $p_\varepsilon^d$. In addition to steady-state data (after long relaxation time), data instantly after compression and transient data between instant and steady states are shown in Fig. 4. Such an unexpected time-dependence of PT kinetics confronts the conventional view that strain-induced PTs do not occur at fixed plastic strain, time is not an essential parameter and plastic strain serves as a time-like parameter, like in Eq. (1) [7, 8, 10]. Note that since the thickness of the sample does not change between instant and steady states, creep as a reason for the time dependence of the strain-induced PT is excluded. It appears that rough-DA allows us not only to reveal the time-dependent part of the growth for strain-induced PT, but also to change the plastic strain-dependent part. As expected, the kinetics for smooth-DA is of the first order, with parameters $a=1$, $k=11.65$, and $B=1.35$ (Fig. S5). Surprisingly, for rough anvils, c-I curve is linear for steady state and instant state before relaxation at 2 GPa step (I < 0.5) and after relaxation, with practically the same slope (Fig. 4b). Thus, the rate of PT in Equation (2) is independent of c, which results in $a=l=0$, $B=1$, and

$$\frac{dc}{dq} = k \frac{p_\alpha(q)-p_\varepsilon^d}{p_h^d-p_\varepsilon^d}. \tag{2}$$

Value $a=1$ corresponds to multiple nucleation within the parent phase, while $a=0$ is typical for propagation from a limited number of nuclei without their interaction, like for thickening of PT band. Equation (2) should be used for each fast-loading increment and for steady state, with different $k$. Time-dependent contribution to the kinetics that reproduces Eq. (2) for the instant kinetics at $t = 0$ and steady-state kinetics for $t = \infty$ and describes transient data at 2 GPa step is:

$$c(t) = c(q)_{t=\infty} + (c(q)_{t=0} - c(q)_{t=\infty}) \, exp\left(-\frac{t}{43.13}\right) \tag{3}$$

with a characteristic time of 43.13 min. Here, $c(q)_{t=0}$ and $c(q)_{t=\infty}$ are the volume fractions after instant compression and in the steady state. Possible rationale for the time dependence is presented in [17] by revisiting results of the phase-field simulations [25, 26] (Fig. S6). It was generally assumed that the time from barrierless nucleation to reaching the equilibrium configuration for each strain increment is on the order of seconds and cannot be detected in experiments. We found this time is ~1 hour, and we could resolve it in our in-situ measurements. This discovery, which could be done by in-situ study only, opens unexplored field of the simultaneous strain- and stress-induced PTs under pressure.

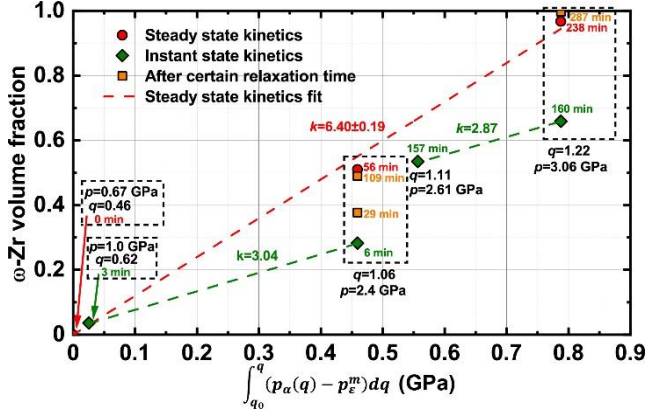

**Fig. 4**. The volume fraction of ω-Zr vs. plastic strain and time. Green diamonds represent diffraction data after instant compression; red circles designate the data used for constraining steady-state kinetics; time labels show time from the beginning of the first measurement; $q$-values are shown near symbols. Revealed zero-order linear strain-dependent kinetics and time dependence of the kinetics of strain-induced PT were not observed in previous literatures.

*Concluding remarks.* The main results summarized at the end of the Introduction not only present the basic and very nontrivial rules of coupled plastic strain-induced PT and nanostructure evolution during severe plastic strain but also open new windows for utilizing rough-DA and finding similar laws for multiple material systems in a broad pressure range and transform the main challenge in DAC experiments—very nonuniform fields—into a prominent opportunity. In addition to enriching an understanding of fundamental material science, these results have significant application potential. Instead of severe plastic straining by high-pressure torsion of a large-grain material (5-6 GPa and 5 anvil turn for Zr [4, 22]), one can reach one of the steady nanostructures by severe straining at normal pressure (e.g., by equal channel or other types of extrusion or rolling) in a much larger sample. Then one can produce partial or complete PT to stronger ω-Zr with smaller grain size by compression up to 3 GPa at relatively small plastic strain or even at lower pressure by high-pressure torsion. For a small volume fraction of ω-Zr, crystallite size is much smaller, and dislocation density is larger than that after complete PT. This gives an idea of designing α-ω Zr composites with increased strength due to strong ultrafine-grained ω-Zr and sufficient plasticity due to ductile α-Zr, and optimizing parameters using Fig. 3. During intense loading, an increase in volume fraction of ω-Zr leads to energy absorption and an increase in strength. All these may result in the economic plastic strain-induced synthesis of nanostructured high-strength high-pressure phases at low pressures. Discovered time-dependence of the kinetics of strain-induced PTs opens an unexplored field of the simultaneous strain- and stress-induced PTs under pressure and is useful for promoting PT at constant load. By optimizing anvil asperities, desirable plastic flow, minimum grain size, and minimum PT pressure can be reached. With rough-DA, friction stress is expected to reach the yield strength in shear $\tau_y$. This will allow one to: (a) determine the pressure-dependence of the yield strength for various material systems using the pressure-gradient method [13, 14]; (b) provide robust boundary conditions for friction for simulating deformation-PT processes in DAC [18, 19] and rotational DAC [8, 30, 31]; (c) increase the maximum possible pressure in DAC without producing toroidal grooves [32], and (d) in-situ study in rotational DAC [3, 6, 8] with maximum friction reaching $\tau_y$, like the traditional high-pressure torsion with ceramic/metallic anvils.

**Acknowledgements**

The authors thank (a) Drs. Alexander Zhilyaev and María-Teresa Pérez-Prado for providing Zr sample; (b) Dr. Reinhard Boehler for preparing the surface of rough-DA. Support from NSF (CMMI-1943710) is greatly appreciated. This work is performed at HPCAT, Advanced Photon Source, Argonne National Laboratory. HPCAT operations are supported by DOE-NNSA's Office of Experimental Science. The Advanced Photon Source is a U.S. Department of Energy (DOE) Office of Science User Facility operated for the DOE Office of Science by Argonne National Laboratory under Contract DE-AC02-06CH11357.

# Supplementary Materials

# Rules of plastic strain-induced phase transformations and nanostructure evolution under high-pressure and severe plastic flow

*Feng Lin\*, Valery I. Levitas\*, Krishan K. Pandey, Sorb Yesudhas, and Changyong Park*

**Experimental Methods and Details**

A commercially pure (99.8%) α-Zr (Fe: 330 ppm; Mn: 27 ppm; Hf: 452 ppm; S: <550 ppm; Nd: <500 ppm, the same as was used in [23]) slab of initial thickness 5.25 mm was cold-rolled multiple times down to a foil of 163 μm to have a heavily pre-deformed sample with hardness saturation. Three different types of DAC compression experiment (hydrostatic, nonhydrostatic with rough-DA, and nonhydrostatic with traditional smooth anvils) were performed. Hydrostatic compression was performed on specks of ~20 μm chipped off from the foil to determine the PT initiation pressure under hydrostatic loading $p_h^d$. 3 mm disks were punched out from thin foil for unconstrained non-hydrostatic compression experiments with both rough-DA and smooth anvil for additional comparison. In-situ axial XRD were performed at Sector 16-BM-D, Advanced Photon Source (APS), Argonne National Laboratory with an x-ray of wavelength 0.3100 Å recorded with Perkin Elmer detector. For rough-DA experiment, diffractions were taken along two perpendicular culet diameters (230 μm) in 10 μm step size. For the smooth anvil experiment, the sample was scanned along one culet diameter (500 μm) in 10 μm step size. Pressure is calibrated using 3rd Birch-Murnaghan equation of state of Zr from [9]. The sample thickness $h$ (see Table S1) was measured by x-ray intensity absorption using the linear attenuation equation with density corrected to the corresponding pressure, similar to [9]. The diffraction images were first converted to unrolled patterns using FIT2D software [33] and then analyzed through Rietveld refinement using Material Analysis Using Diffraction (MAUD) software [34] to obtain the lattice parameters, volume fractions of ω-Zr, microstrains, and crystallite sizes.

The dislocation density is estimated with the crystallite size and microstrain using Williamson-Smallman method [35]:

$$\rho = \sqrt{\rho_c \rho_{ms}} \; ; \qquad \rho_c = \frac{3}{d^2} \; ; \quad \rho_{ms} = k\varepsilon^2/b^2. \tag{S1}$$

Where $\rho_c$ and $\rho_{ms}$ are the contribution to overall dislocation density from crystallite size and microstrain, respectively. $d$ is the crystallite size and $\varepsilon$ is the microstrain; $b$ is the magnitude of the Burgers vector; $k = 6\pi A(\frac{E}{G \ln(r/r_0)})$ is a material constant; $E$ and $G$ are Young's modulus and shear modulus respectively; $A$ is a constant that lies between 2 and $\pi/2$ based on the distribution of strain; $r$ is the radius of crystallite with dislocations; $r_0$ is a chosen integration limit for dislocation core. In this study, $A = \pi/2$ as the gaussian distribution of strain. Moduli $E$, $G$ and their pressure dependence for α and ω-Zr are taken from [36] and [37], respectively. A reasonable value of $\ln(r/r_0)$ being 4 is used [35]. Burger vector is extracted from the dominate slip system. α-Zr has a dominant prismatic slip system of $\{1\bar{1}00\}\langle11\bar{2}0\rangle$ [38-41]. As for ω-Zr, a prismatic $\{11\bar{2}0\}\langle1\bar{1}00\rangle$ and basal $\{0001\}\langle1\bar{1}00\rangle$ dominant slip system is suggested based on plasticity modeling [42].

**Table S1.** The thickness of Zr sample with rough-DA in this study at corresponding compression step. 0.67 GPa corresponds to the step when ω-Zr first emerge at culet center.

| Compression step | initial | 0.67 GPa | 2 GPa | 3 GPa |
|---|---|---|---|---|
| Thickness (μm) | 163 | 101 | 56 | 48 |

## Supplementary Discussion

**1. Rationales for the evolution of the crystallite size and dislocation density in ω-Zr during the phase transformation**

Small crystallite size in ω-Zr at the beginning of PT is caused by small transformed regions. The growth of the crystallite size in ω-Zr is related to the growth of these regions in the course of PT. Also, as it follows from [9] and the current paper, the reduction in the crystallite size of α-Zr reduces the minimum pressure for initiation of the strain-induced PT $p_\varepsilon^d$ and promotes the PT. That is why the smallest crystallites of α-Zr transform first to ω-Zr, then larger grains transform, so the crystallite size in ω-Zr grows during PT. Since ω-Zr is approximately two times stronger than α-Zr, plastic strain is mostly localized in the α-Zr. That is why plastic strain and strain path do not affect the crystallite size and dislocation density in ω-Zr. Reduction in the dislocation density in ω-Zr is caused by the inverse proportion between the dislocation density and the crystallite size following from Eq. S1.

**2. Explanation of the existence of outliers in the evolution of the crystallite size and dislocation density in α-Zr during the phase transformation**

As it follows from Fig. 3, the crystallite size of and dislocation density in ω-Zr during the phase transformation are unique functions of the volume fraction of ω-Zr independent of pressure, plastic strain tensor, and its path. Similar dependence is found for α-Zr, but there are outliers for 0.4<$c$<0.5 obtained at the 2 GPa step. Indeed, at the 2 GPa step and in the two-phase region, the crystallite size of α-Zr remains constant while its volume fraction is larger than 0.6 (Fig. 2), same as the steady value before PT. When the volume fraction of α-Zr gradually decreases to 0.5 towards the culet center, the average crystallite size of α-Zr slightly increases to ~60 nm. This is caused by the statistical effect. As it follows from [9] and the current paper, the reduction in the crystallite size of α-Zr reduces the minimum pressure for initiation of the strain-induced PT $p_\varepsilon^d$ and promotes the PT. Smallest crystallites of α-Zr transform first to ω-Zr, increasing the average size of the remaining α-Zr crystallites. It is almost non-detectable at large volume fractions of α-Zr but essential at small volume fractions. Also, constant crystallite size is observed for $r$>60 μm, whereplastic deformation is much larger than at the central part due to friction,. This large plastic strain restores the same steady averaged crystallite size by refining large crystallites. At the center, plastic strain is much smaller and insufficient to restore the steady size. At the 3 GPa step, with further reduction in the volume fractions of α-Zr and an increase in plastic strain, these outliers disappear, and all points belong to the single red curve in Fig. 3 versus volume fractions of ω-Zr. Reduction in crystallite size is related to dividing α-Zr crystallite into two or more parts due to PT inside of grains.

A similar statistical effect can explain outliers in the dislocation density in α-Zr for 0.4<$c$<0.5 obtained at 2 GPa step. Formally, it is caused by the inverse proportion between the dislocation density and the crystallite size that follows from Eq. S1. Physically, PT starts and occurs first in the grains with the largest dislocation density, where the probability of strong stress concentrators is higher. Transformation of these grains of α-Zr to ω-Zr decreases the averaged dislocation density in the remaining α-Zr crystallites. It is almost non-detectable at large volume fractions of α-Zr but essential at decreasing volume fractions. Also, constant dislocation density is observed for $r>60$ μm, where plastic deformation is much larger than at the central part. This large plastic strain restores the same steady averaged dislocation density in the large grains. At the center, plastic strain is much smaller and insufficient for restoring the steady dislocation density. At the 3 GPa step, with further reduction in the volume fraction of α-Zr and an increase in plastic strain, these outliers disappear, and all points belong to the single red curve in Fig. 3 versus the volume fraction of ω-Zr. An increase in averaged dislocation density is probably caused by increased dislocation density near new α-ω interfaces to accommodate local transformation strain and decreased crystallite size. Large scatter in both crystallite size and dislocation density in α-Zr near completion of PT is caused by increasing measurement error for a tiny amount of α-Zr.

**3. On the possible source of the time dependence of the kinetics of strain-induced PTs**

It was generally accepted that during shear under high pressure, PT stops when shear stops [1,2,7-10]. That means that time is not a governing parameter and plastic strain plays a role of a time-like parameter. A nanoscale rationale in [7] explaining this statement was that barrierless nucleation at the tip of the dislocation pileup occurs extremely fast. Since stress decreases like $1/r$ with distance from the tip $r$, grows is very limited and is arrested when phase interface is equilibrated. This process is assumed to occur in a much shorter time than the measurement time, a time-dependent component is not detectable, and plastic strain is the only governing parameter. This was implemented in [7] in the strain-controlled kinetic equation, see Eq. (1) in the main text. This equation was confirmed by experiments in [9], but the time-dependent component of the kinetics at fixed load/torque was not checked because it was not expected. After we found here the time dependence of the kinetics of strain-induced PT experimentally, we can revisit the results of the phase-field simulations to rationalize it. Fig. S6 shows the time evolution of the phase and dislocation structures at the fixed applied normal stress and shear strain after phase nucleation in the right grain at the dislocation pileup in the left grain. Applied normal stress is 10 times lower than the PT pressure under hydrostatic conditions. One can see the following sequence of events after nucleation: (1) the high-pressure phase significantly grows and reaches the opposite grain boundary;
(2) the number of dislocations in the dislocation pileup in the left grain increases (especially within step at the grain boundary);
(3) dislocations nucleate and evolve in the right grain;
(4) the second nucleus appears at the dislocation pileup that develops within the right grain;
(5) nuclei coalesce, and the stationary phase and dislocation configurations is achieved.

The time scale for phase and dislocation evolution is determined by two corresponding kinetic coefficients, which are different for different materials. If the first measurement at the material point in a sample in DAC completes after a stationary state is reached, this evolution is undetectable, the entire process looks instantaneous, and the kinetics of the PT is fully plastic strain controlled. In the opposite case, phase evolution at the fixed strain will be observed and time-dependent component of the kinetics should be characterized and formalized. Since postmortem studies require time for pressure release, sample preparation, and measurements, it is not surprising that the time-dependent kinetics was not discovered before.

**Supplementary Figures**

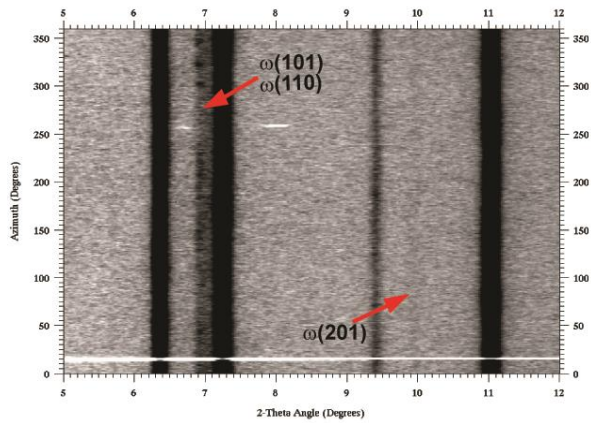

**Fig. S1. Unrolled diffraction image of Zr when ω-Zr first emerged at 0.67 GPa at culet center.**

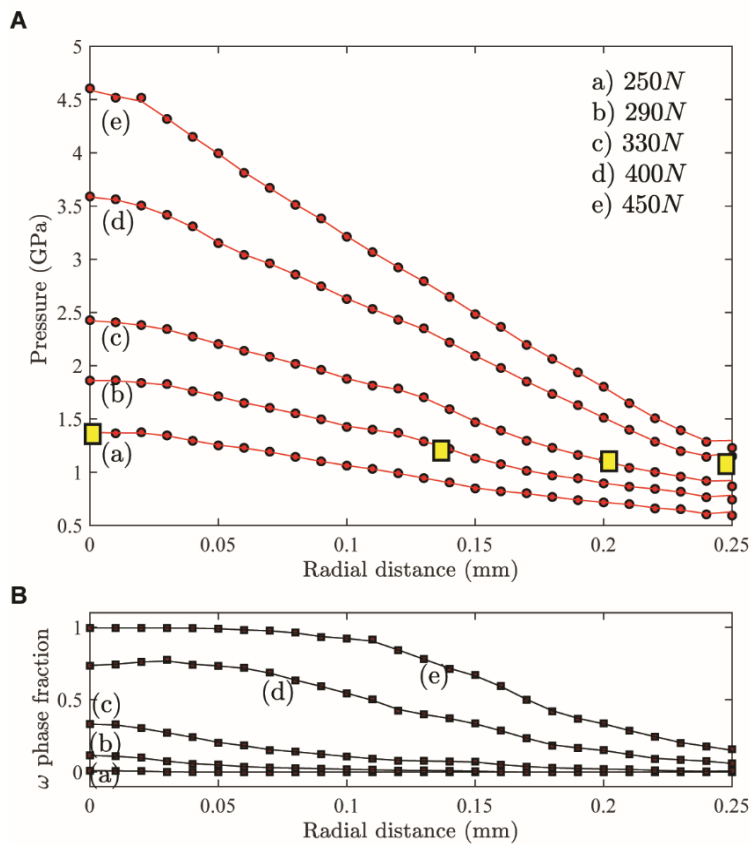

**Fig. S2. Radial distribution of (A) α-Zr pressure and (B) ω-Zr volume fraction in a sample deformed with smooth anvils.** Different applied forces represent different compression stages. Yellow squares show the minimum PT pressure $p_\varepsilon^d$ =1.36 GPa at different compression stages and at different radii where ω-Zr was first observed. Since plastic strain, plastic strain path, and pressure-strain path are very different at different locations and compression stages and $p_\varepsilon^d$ is independent of the locations, then $p_\varepsilon^d$ is independent of plastic strain, plastic strain path, and pressure-strain path.

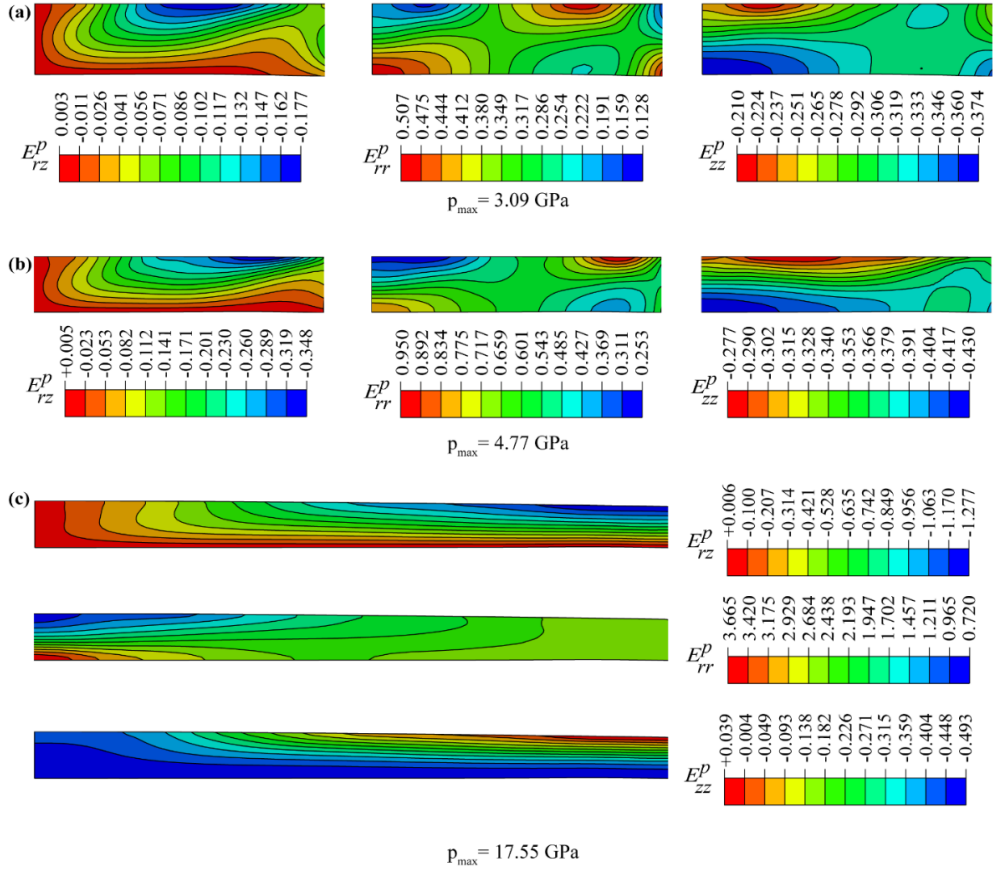

**Fig. S3. Distributions of components of Lagrangian plastic strains in a quarter of a sample for three loadings charcterized by the maximum pressure in a sample.** Very heterogeneous and nontrivial distributions are observed, caused by heterogeneous contact friction. At the symmetry axis (left side of a sample) and symmetry plane (bottom of a sample), shear strains $E^p_{rz}$ are zero. At the contact surface with a diamond (top of a sample) shear strains and particle rotations reach their maximum due to large contact friction. During compression, each material particle flows radially in the reagion with larger shear and different proportion of the normal strain, i.e., is subjected to complex nonproportional straining, very different from othe particles. Thus, numerous plastic strain tensors and straining paths are realized. Adopted with changes from [20] with permissions.

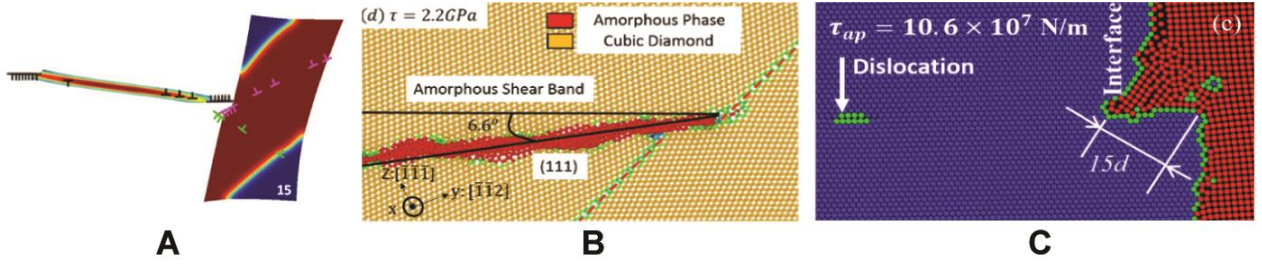

**Fig. S4. Dislocation pileups produce a step at the grain boundary or phase interface that causes a phase transformation.** (**A**) Dislocation pileup in the left grain produces step at the grain boundary and cubic to tetragonal PT and dislocation slip in the right grain. Phase-field approach results from [25]. (**B**) Dislocation pileup in the right grain produces a step at the grain boundary in Si I and amorphization in the left grain. Molecular dynamics results from [28]. (**C**) Step at the phase interface boundary consisting of 15 dislocations and causing cubic to hexagonal PT. The atomistic portion of the concurrent continuum-atomistic approach from [52]. Adopted with changes from [25, 28, 29] with permissions.

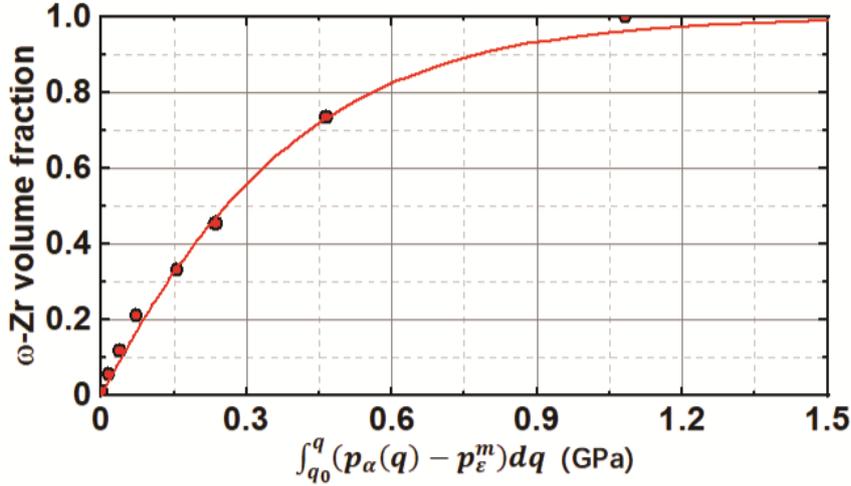

**Fig. S5. Kinetics of α-ω phase transformation in Zr with smooth diamond anvil.** Compared to rough-DA experiment, kinetics shows different nonlinear features corresponding to the first-order reaction with parameters $a=1$, $k=11.65$, and $B=1.35$ in Equation (2) $\frac{dc}{dq} = k\frac{B(1-c)^a}{B(1-c)+c}\left(\frac{p_\alpha(q)-p_\varepsilon^d}{p_h^d-p_\varepsilon^d}\right)$, instead of $a=l=0$, $B=1$ for the experiment with rough-DA.

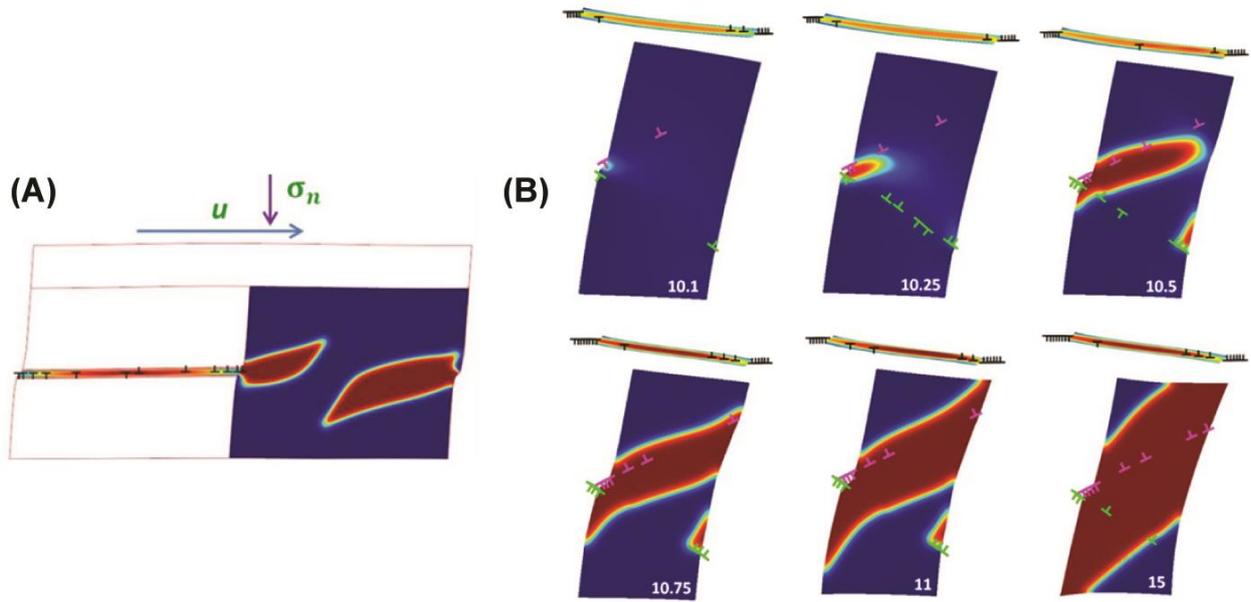

**Fig. S6. Time evolution of the phase and dislocation structures at the fixed applied normal stress and shear strain after phase nucleation in the right grain at the tip of dislocation pileup in the left grain.** (A) Schematics of grains with an initial solution for dislocation pileup and nucleated high-pressure phase (red) [26]. (B) Nucleation and growth of the high-pressure phase (red) in the right grain caused by an evolving dislocation pileup in the left grain, which is shown at the top of each right grain [25]. Results are obtained with the phase-field approach. Adopted with changes from [25, 26] with permission.